\documentclass[11pt]{article}
\usepackage{graphicx}
\usepackage{epsfig}
\usepackage{amssymb,amsfonts,amsmath,amsthm}
\usepackage{mathrsfs}
\usepackage{epstopdf}
\usepackage{hyperref,color}
\usepackage{natbib}
\usepackage{setspace}
\usepackage{authblk}

\newtheorem{defi}{Definition}

\newtheorem{axiom}{Axiom}

\pdfoutput=1

\begin{document}

\title{A quantum-information-theoretic complement to a general-relativistic implementation of a beyond-Turing computer}
\author[$\dag$]{Christian W\"uthrich\thanks{I am grateful to Istv\'an N\'emeti and Hajnal Andr\'eka for discussions on the topic and for their willingness to share old material from their personal archive. I thank Gergely Sz\'ekely for his great patience with my procrastinating, Joseph Berkovitz and Kent Peacock for correspondence, and the audience at the Istv\'an-Fest, as well as John Dougherty and the referees for this journal for comments.}}
\affil[$\dag$]{Department of Philosophy, University of California, San Diego}
\date{11 June 2014}
\maketitle

\begin{abstract}\noindent
There exists a growing literature on the so-called physical Church-Turing thesis in a relativistic spacetime setting. The physical Church-Turing thesis is the conjecture that no computing device that is physically realizable (even in principle) can exceed the computational barriers of a Turing machine. By suggesting a concrete implementation of a beyond-Turing computer in a spacetime setting, Istv\'an N\'emeti and Gyula D\'avid (2006) have shown how an appreciation of the physical Church-Turing thesis necessitates the confluence of mathematical, computational, physical, and indeed cosmological ideas. In this essay, I will honour Istv\'an's seventieth birthday, as well as his longstanding interest in, and his seminal contributions to, this field going back to as early as 1987 by modestly proposing how the concrete implementation in N\'emeti and D\'avid (2006) might be complemented by a quantum-information-theoretic communication protocol between the computing device and the logician who sets the beyond-Turing computer a task such as determining the consistency of Zermelo-Fraenkel set theory. This suggests that even the foundations of quantum theory and, ultimately, quantum gravity may play an important role in determining the validity of the physical Church-Turing thesis. 
\end{abstract}


\begin{center}
{\em To Istv\'an N\'emeti on the occasion of his 70th birthday}
\end{center}

\noindent
The physical Church-Turing thesis, as I will understand it, asserts that no computing device that is physically realizable (even in principle) can exceed the computational barriers of a Turing machine. It is not to be confused with the logically independent Church-Turing thesis, according to which a function is `effectively', or `algorithmically' computable just in case it is computable by a Turing machine.\footnote{And hence, the argument presented in this essay is not touched by the complaints by \citet[94]{shapit03}.} This essay will be concerned with the physical realizability rather than with issues in the foundations of computation theory. 

There are at least two ways in which the locution `physically realizable' can be fleshed out. First, it may mean that an idealized device is realizable or implementable in the context of a specific physical theory. In the present setting, one such specific theory would be general relativity. Alternatively, it may mean that an idealized device is realizable or implementable as judged by all our fundamental theories, or our current best candidates for such theories, or a carefully balanced combination of our current best candidates. Thus, unlike the Church-Turing thesis, or Goldbach's conjecture, the physical Church-Turing thesis is not just a hypothesis in pure mathematics, but crucially involves physics. What kind of physics it involves depends on how we contextualize `physically realizable'. After a very brief history of relativistic computing in \S\ref{sec:hist}, we will start in \S\ref{sec:mala} by confining ourselves to general relativity. In \S\ref{sec:comm}, the purview of our considerations will be extended to include quantum physics. The conclusions follow in \S\ref{sec:conc}. 

At the first pass, the question thus arises as to whether general relativity permits, at least in principle, a scenario in which `hyper-Turing' computations, i.e., computations not executable by Turing machines, are possible. The main thesis of \citet{etenem}, and repeated in \citet{nemdav}, holds that ``it is consistent with Einstein's equations, i.e.\ with general relativity, that by certain kinds of relativistic experiments, future generations might find the answer to non-computable questions like the halting problem of Turing machines or the consistency of Zermelo Fraenkel set theory'' \citep[119]{nemdav} and hence that the physical Church-Turing thesis is false, at least in the context of general relativity. The goal of the present essay is to make modest progress in spinning some open threads of this literature a bit further.

\section{A brief history of relativistic computers}
\label{sec:hist}

Before I delve into the details of the physical implementation of a device that could realize a beyond-Turing computer, let me briefly recount some of the history behind the leading ideas of this literature.\footnote{Cf.\ also \citet[\S3]{nemdav}.} So as we celebrate Istv\'an's 70th birthday in 2012, we can equally hail the 25th anniversary of relativistic computers. As extant lecture notes prove, Istv\'an N\'emeti's 1987 PhD seminar entitled `On logic, relativity, and the limitations of human knowledge' at Iowa State University first articulates the idea of a relativistic computer as it will essentially be given in Definition \ref{def:relcomp} below. Shortly thereafter, in 1988, David Malament privately communicates the idea of what we will call a `Malament-Hogarth spacetime' below in Definition \ref{def:mhspt} to John Earman. In 1990, Itamar Pitowsky develops the related idea of what was coined in \citet{earnor93}---another seminal early contribution---`Pitowsky spacetime' \citep{pit90}. Recognizing the significance of Malament's general ideas about requisite spacetime structures, Mark Hogarth generalizes them and applies them to computability theory \citep{hog92}.  While the succession of novel contributions to the literature remains rapid today, these early ideas collectively establish the field of relativistic computers within five years of its inception.

\section{A Malament-Hogarth spacetime: Kerr-Newman}
\label{sec:mala}

Any assessment of the {\em physical} Church-Turing thesis will thus depend, at least in part, on {\em physical} possibility as legislated by {\em physical} theories. Clearly, such an evaluation must be grounded in the best available physical theories for it to claim any validity. Potentially, this includes distinct and quite different theories. First, as is recognized by many, general relativity (GR), as our best spacetime theory to date, offers novel and sometimes surprising spatiotemporal settings for computations, and may thus permit beyond-Turing computers. To elaborate how this possibility arises is the goal of this section. Before we embark, however, let me emphasize that other physical theories may well be considered in this context; after all, a computational device will presumably be built from material components, and GR tells us nothing about most of the properties of matter and of energy. Thus, it wouldn't be all that surprising if some quantum theory or other became relevant. In fact, as we shall see in \S\ref{sec:comm}, quantum theories take centre stage in my attempts to codify a communication protocol between the computer and its programmer. Finally, it should be noted that we must reasonably expect that many of the issues we will encounter will only find a final resolution in a full quantum theory of gravity. In the absence of such a theory, however, let us proceed for now with the best available (classical) spacetime theory---GR.

A model in GR is a triple $\langle \mathcal{M}, g_{ab}, T_{ab}\rangle$ such that the Einstein field equations are satisfied, where $\mathcal{M}$ is a four-dimensional Lorentzian manifold, $g_{ab}$ the metric tensor, and $T_{ab}$ the energy-momentum tensor. An ordered pair $\langle\mathcal{M}, g_{ab}\rangle$ such that there exists a $T_{ab}$ for which the triple $\langle \mathcal{M}, g_{ab}, T_{ab}\rangle$ is a model of GR is called a {\em relativistic spacetime}. As it turns out, there are relativistic spacetimes which permit the implementation of a beyond-Turing computer. In GR, physical clocks `deeper' in a gravitational well, i.e., at spacetime locations with stronger gravitational fields, run more slowly than their initially synchronized counterparts `higher up' in the gravitational field. This differential in field strength can now be used to achieve a computational speed-up. In fact, as there is no upper bound on the gravitational field strength in GR, the achievable speed-up is also beyond any bounds. 

\begin{figure}
\centering
\epsfig{figure=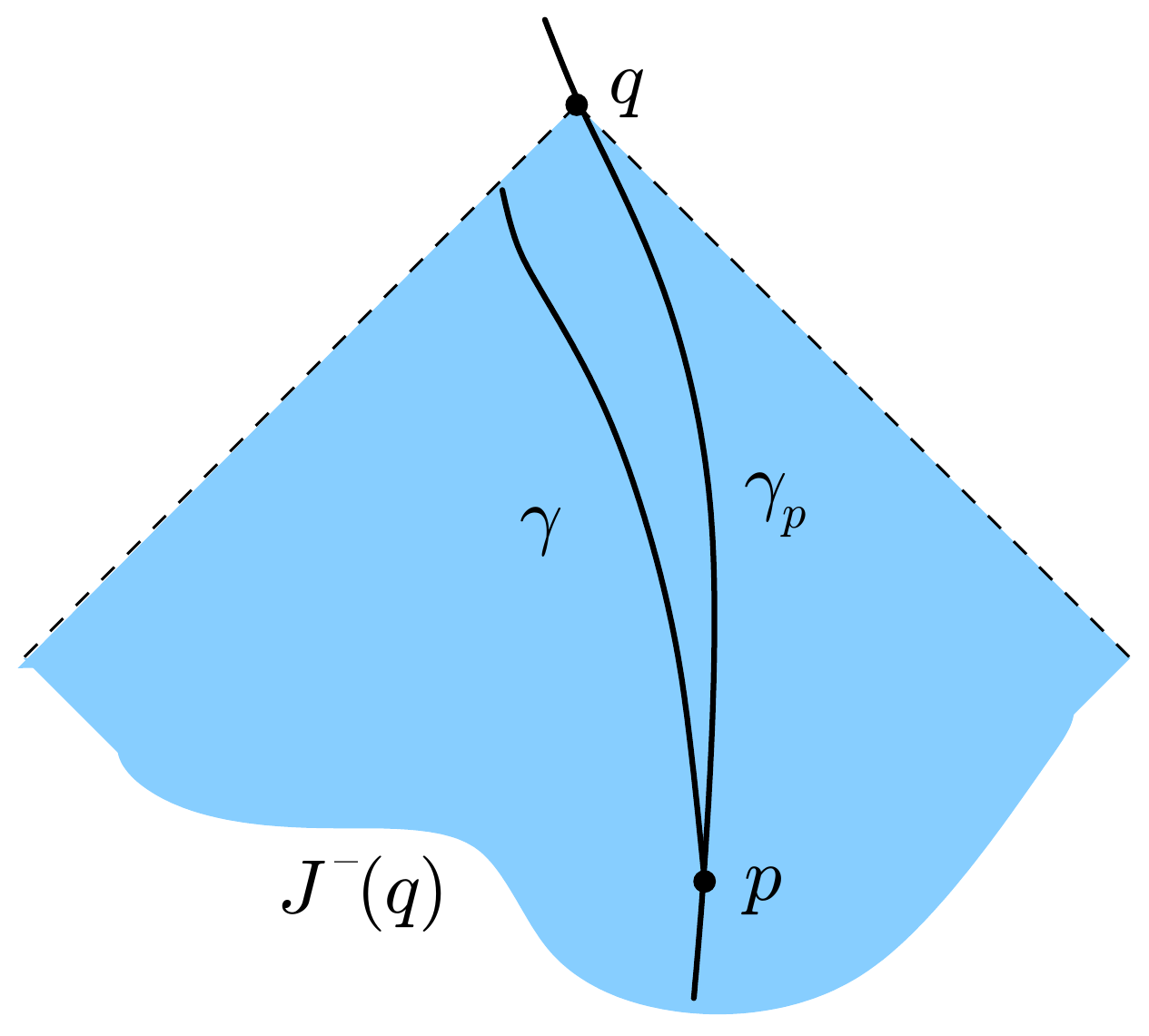,width=0.6\linewidth}
\caption{\label{fig:mhspt} A Malament-Hogarth spacetime with the region $J^-(q)$ shaded in blue.}
\end{figure}

In order to obtain beyond-Turing computation, the gravitational field must be such that the computer can run an infinite number of computations in the causal past of some spacetime event the programmer can reach within finite time. It is essential to appreciate the fact that this hypercomputational design does not presume an {\em intrinsic} speed-up of the computer; in particular, it does not rely on the feasibility of a `supertask', at least not if this locution is taken to imply the possibility of the performance or operation of an infinite number of tasks within a finite amount of proper time {\em of the device which does the computing}. Instead, what relativistic hypercomputation relies on is the existence of a worldline $\gamma$---the computer's worldline---with infinite proper length,
\[
\int_\gamma ds = \infty,
\]
which completely remains in the causal past of an event which can be reached by the programmer in finite proper time along {\em her} worldline. Relativistic spacetimes in which such a worldline exists are called {\em Malament-Hogarth spacetimes}:
\begin{defi}\label{def:mhspt}
A relativistic spacetime $\langle\mathcal{M}, g_{ab}\rangle$ is a {\em Malament-Hogarth spacetime} just in case there is a future-directed timelike half-curve $\gamma: \mathbb{R}^+ \rightarrow \mathcal{M}$ such that $\int_\gamma ds = \infty$ and there exists a point $q\in\mathcal{M}$ (the {\em MH event}) satisfying im$\gamma \subset J^-(q)$, where $J^-(q)$ is the causal past of $q$.
\end{defi}
The salient aspect of a Malament-Hogarth spacetime is illustrated in Figure \ref{fig:mhspt}, where the entirety of the curve $\gamma$ is confined to the causal past $J^-(q)$ of an event $q$ on the worldline $\gamma_p$ of a programmer. The computer's worldline $\gamma$ and the programmer's worldline $\gamma_p$ part ways at $p$.
A certain kind of determinism fails in Malament-Hogarth spacetimes, as they are not globally hyperbolic and thus do not admit a Cauchy surface.\footnote{A {\em Cauchy surface} is a spacelike hypersurface $\Sigma$ such that every timelike or null curve without endpoints intersects $\Sigma$ exactly once.} It is straightforward to see that this must be the case. A globally hyperbolic spacetime can be foliated by totally ordered Cauchy surfaces such that the topology of the spacetime manifold is $\mathbb{R}\times \Sigma$, where $\Sigma$ denotes any Cauchy surface.\footnote{\citet{wald1984}, Theorem 8.3.14.} If a Malament-Hogarth spacetime were globally hyperbolic, the Malament-Hogarth event $q$ would thus be contained in some Cauchy surface. Thus, every inextendible timelike curve would have to intersect it. But this is clearly impossible for the half curve $\gamma$, which cannot be extended to leave the causal past of $q$. Thus, Malament-Hogarth spacetimes are not globally hyperbolic.

A Malament-Hogarth spacetime is required to built a `relativistic computer', i.e., a hypercomputational device or set-up of devices, which relies on the spacetime structure of a relativistic spacetime to achieve beyond-Turing computation, rather than on an intrinsic speed-up. Following \citet[Def.~4.5]{nemdav}, we define a {\em relativistic computer} quite abstractly as follows:
\begin{defi}[Relativistic computer]\label{def:relcomp}
A {\em relativistic computer} in a Malament-Hogarth spacetime $\langle\mathcal{M}, g_{ab}\rangle$ is a triple $\langle \gamma_p, \gamma, q\rangle$ such that $\gamma$ is an upward-infinite future-directed curve fully in $J^-(q)$ of some event $q\in\mathcal{M}$, $\gamma_p$ is a timelike curve such that $q$ lies on it and an initial segment of $\gamma_p$ coincides with $\gamma$ (up to, say, point $p$).
\end{defi}
Such a device should surely qualify as a hypercomputational device, as there is a clear sense in which the programmer will, in principle, be able to witness the computer perform an infinity of tasks within a finite amount of the programmer's own proper time. Since we are interested in the {\em physical} Church-Turing thesis, the questions before us are (i) whether GR permits Malament-Hogarth spacetimes such that a relativistic computer can be concretely implemented, and (ii) if so, whether any of these concrete implementations are at all physically reasonable. 

\begin{figure}
\centering
\epsfig{figure=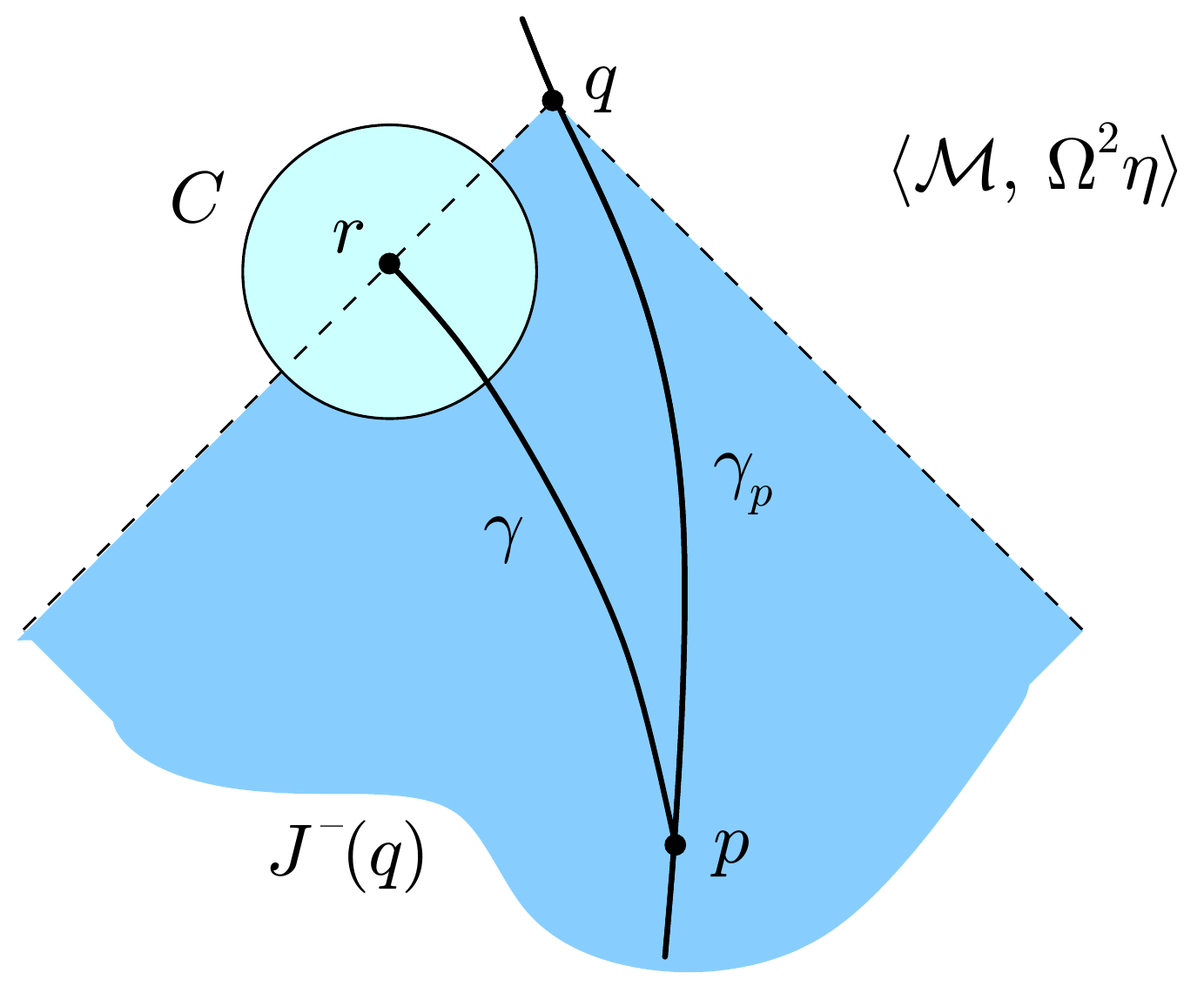,width=0.6\linewidth}
\caption{\label{fig:minkasmala} The conformally `blown up' Minkowski spacetime.}
\end{figure}

It turns out that there are many relativistic spacetimes which permit the operation of relativistic computers. For instance, if one takes Minkowski spacetime $\langle \mathbb{R}^4, \eta_{ab}\rangle$, punches a hole in it, and conformally `blows up' a compact region around the punctuation, one obtains such a spacetime. As illustrated in Figure \ref{fig:minkasmala}, the metric field of the original Minkowski spacetime is modified by a conformal factor $\Omega^2$, which is the identity outside of a compact region $C$, but goes to infinity as event $r$, which is removed from $\mathbb{R}^4$, is approached. The spacetime $\langle \mathbb{R}^4\setminus \{r\}, \Omega^2\eta_{ab}\rangle$ is a Malament-Hogarth spacetime because $\int_\gamma ds = \infty$ (and all other conditions are also satisfied).

Another example is anti-de Sitter spacetime, a Lorentzian analogue of a hyperbolic space with constant negative scalar curvature. The most important family of Malament-Hogarth spacetimes for our present purposes, however, are the Kerr-Newman spacetimes which describe rotating, charged black holes. Such a black hole is classically defined by three parameters: the black hole's mass $M$, its angular momentum $a$, and its electric charge $Q$. Kerr-Newman spacetimes are Malament-Hogarth spacetimes if they satisfy the condition $|a| + |Q| \neq 0$. Kerr-Newman spacetimes include the special cases of Kerr spacetimes, which describe vacuum spacetimes of uncharged, rotating black holes (and are Malament-Hogarth spacetimes if $|a| < M$), and of Reissner-Nordstr\"om spacetimes, which represent charged, but non-rotating black holes. In what follows, we shall consider a Kerr-Newman spacetime of a slowly rotating, slightly charged black hole.

\begin{figure}
\centering
\epsfig{figure=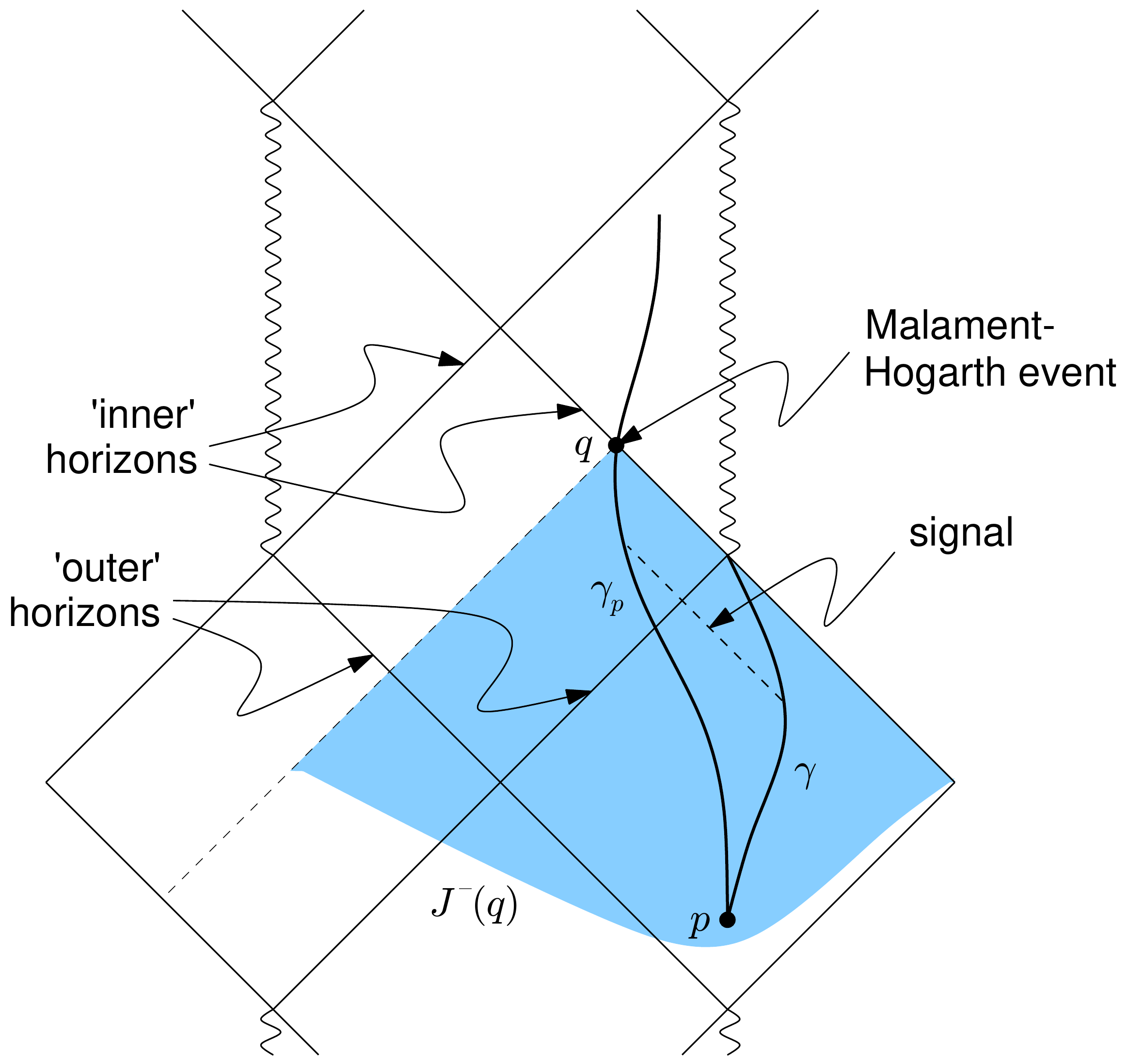,width=0.8\linewidth}
\caption{\label{fig:kerrnewman_penrose} Penrose diagram of an analytically extended Kerr-Newman spacetime, with the worldline of a signal between computer and programmer shown.}
\end{figure}

The causal structure of a Kerr-Newman spacetime is given by its Penrose diagram, as in Figure \ref{fig:kerrnewman_penrose}, with the timelike singularities indicated by curly vertical lines. These diagrams capture the conformal structure of a spacetime, which makes it straightforward to determine whether the represented spacetime is a Malament-Hogarth spacetime. As can be gleaned from Figure \ref{fig:kerrnewman_penrose}, all the elements of a Malament-Hogarth spacetime and thus of a relativistic computer are in place in a(n analytically extended) Kerr-Newman spacetime.\footnote{For a more detailed treatment of Kerr-Newman spacetime in the context of relativistic computation, cf.\ \citet[\S2]{nemdav}; for an accessible entry to the main ideas, cf.\ \citet{nemand06}. For systematic treatments of Kerr spacetimes in GR, cf.\ \citet{one95} and \citet{wilvissco}.}
The proper length of the programmer's worldline $\gamma_p$ is finite, while that of the computer $\gamma$ is infinite, and all of $\gamma$ lies in the causal past $J^-(q)$ of $q$. As the figure illustrates, the programmer must dive past the black hole's horizon in order to have the entirety of $\gamma$ in her past and thus to reap the benefits of the relativistic computation. In fact, the Malament-Hogarth event $q$ lies on the inner horizon of the Kerr-Newman black hole. All black holes possess an `outer' event horizon, i.e., a hypersurface enveloping the spacetime region from which neither massive bodies nor light can escape the gravitational pull of the black hole. Kerr-Newman spacetimes also have `inner' event horizons, due to the repelling effect of the centrifugal force and occasionally of the electrostatic force. Such an inner horizon forms if this repelling force overcomes the gravitational attraction of the mass of the black hole and thus marks the hypersurface where both forces counterbalance one another. 

If the programmer thus finds herself in such a Kerr-Newman spacetime, she can take advantage of its Malament-Hogarth character and perform a beyond-Turing computation. A question known to be non-Turing computable is, for instance, the consistency of Zermelo-Fraenkel set theory with the Axiom of Choice ({\sf ZFC}), i.e., whether we can derive the formula {\sf FALSE} from the axioms of {\sf ZFC}.\footnote{Cf.\ \citet[\S12]{earnor96} and \citet{wel08} for an assessment of the extent of beyond-Turing computation in Malament-Hogarth spacetimes. Welch shows that although there are upper bounds to what can be computed (under fairly mild assumptions), effectively Borel statements can be resolved.} The computer can thus check the consistency of {\sf ZFC} by checking propositions one by one---perhaps one a day---to see whether they are theorems of {\sf ZFC}. Obviously, if the computer finds that some proposition $\phi$ and its negation $\neg\phi$ are both theorems of {\sf ZFC}, and thus finds a proof of the formula {\sf FALSE}, then the inconsistency is established, the question answered, and the computer sends a predetermined signal to the programmer and halts. However, as long as the computer does not find such an inconsistency, it continues checking and no signal is sent. Obviously, this set-up assumes, quite unrealistically, that the computer never crashes, is free of programming bugs, has access to an arbitrary large memory and to an arbitrarily large amount of energy, and never encounters any other obstacle to its proper functioning. Alas, none of this is true of the computer on which this essay was written. However, these assumptions are consistent with the laws of physics as best we currently know them and shall thus not stall us in our considerations of matters of principle.\footnote{Cf.\ \citet[\S4.8]{ear95} and \citet[\S5]{nemdav} for an assessment of these potential obstacles. Cf.\ also the earlier \citet{earnor93}, and the recent \citet{man10}. I concur with Manchak's assessment that the physical reasonableness of Malament-Hogarth spacetimes is yet to be settled. As to whether this means that a {\em relativistic computer} is physically reasonable, cf.\ \citet[\S4.2]{pic11}. Piccinini accepts that relativistic computers are in principle possible (although computations in our Kerr-Newman computer are not {\em repeatable}), but nevertheless concludes that ``for now and the foreseeable future, relativistic hypercomputers do not falsify [the physical Church-Turing thesis].'' (759) On the question of repeatability, cf.\ also \citet[\S5]{andeal09}.} 

Even though it is clear that the causal structure of a Kerr-Newman---courtesy of the non-vanishing angular momentum and charge of the black hole---permits the programmer to avoid collision with the singularity depicted in Figure \ref{fig:kerrnewman_penrose} by the wavy lines, one might worry that the strong gravitational field of the black hole pose a health hazard to our programmer. In particular, strong tidal forces may `spaghettify'\footnote{This is a technical term describing the radial stretching and the tangential squeezing acting on our traveller.} the intrepid programmer. However, as it turns out, these tidal forces become the smaller the bigger the black hole is, making the plunge entirely safe for sufficiently large black holes.\footnote{Cf.\ \citet[\S3.4]{wut99} for a precise calculation of the size of the tidal forces encountered by such a traveller.}

Having put these petty worries aside, the decisive question is how the programmer can {\em learn} whether {\sf ZFC} is consistent. If she received the computer's signal, then she will know that {\sf ZFC} is inconsistent. If, on the other hand, she reaches the MH event without having received a signal, then she will know that {\sf ZFC} is consistent. This means that she can only come to know that {\sf ZFC} is consistent if she can know when she passed the MH event on her trip inside the black hole. This requires the programmer's ability to locate herself in the spacetime, e.g.\ by measuring the local gravitational field. For the sake of argument, let us assume that she has the technological know-how and equipment to perform any local experiment she desires and that she is fully apprised of everything in her causal past. 

Unfortunately, this is not strictly enough for her to determine her future, including how long it will take her to traverse the horizon. Since the horizons in relativistic spacetime are a matter of the {\em global} structure of spacetime, they come without {\em local}, intrinsic, markers that would permit a traveller to `see' the horizon as it is passed. In order for her to know when she passes the MH event, therefore, the programmer needs to have sufficient information about the global structure of the spacetime she inhabits. And the problem is that as an inhabitant of the spacetime with no recourse to a `God-like' external point of view she can never have complete knowledge of the global structure of her spacetime. In fact, under rather mild assumptions, no amount of observations an epistemic agent can have uniquely determines the global structure of spacetime \citep{man09}. This entails that not only can the programmer never be certain that she in fact inhabits a Kerr-Newman spacetime, but she can seemingly never determine the point of passing the inner horizon and thus the MH event.

While the programmer cannot {\em know} that she in fact finds herself in a spacetime appropriately Malament-Hogarth, let us assume that she in fact does live in a Kerr-Newman spacetime such that all spacetime elements requisite for our purposes are in fact present. She will still need to have the means to determine whether she passed the MH event. One seemingly simple way to do so would be to invoke the fact that as she passes the MH event, she will no longer receive light signals from her home galaxy. The night sky will, as it were, turn dark as she travels past the MH event $q$ because all worldlines of objects not diving into the outer horizon of the black hole are confined to $J^-(q)$, as can easily be gleaned from Figure \ref{fig:kerrnewman_penrose}. Careful scrutiny, however, reveals that what appears to be an elegant solution to our problem is in fact accompanied by a fatal effect: if the radiation emanating from any source approaching the future timelike infinity of the spacetime `tile' that contains event $p$ does not peter off at a sufficient rate, then the programmer who receives this radiation in ever shorter intervals of her proper time will be fried to death by the unbounded energy of the incoming radiation. This is the so-called `blueshift problem' noted in the literature before and treated, for individual signal pulses from the computer but applicable to all sources sending radiation towards the programmer, in \citet[Lemma 4.2 and surrounding discussion]{ear95} and \citet[Proposition 6]{etenem}. The only way for the radiation from her neck of the spacetime woods not to kill her is for the emanated radiation to decrease sufficiently rapidly along the worldlines of the sources. 

One might think that the problem can be circumvented by letting the computer send regular digest messages even if it did not yet derive the formula {\sf FALSE}. The rate at which the programmer receives these messages would let her anticipate how far away from the MH event she still was. No such messages would reach her after she passed the MH event, of course, and she would thus know that {\sf ZFC} was consistent. But for the reasons given by \citet[Lemma 4.2]{ear95} and \citet[Proposition 6]{etenem}, a similar blueshift problem would arise in this case and the energy of the light pulses would again roast our poor programmer. This was why we set up the communication protocol between computer and programmer such that the computer only sends a message if it derives the formula {\sf FALSE} and none otherwise. This makes sure that the signals from the computer do not bake the programmer, but it does not help her determine whether she passed the MH event. 

Apart from the issues involving the communication between computer and programmer addressed in \S\ref{sec:comm}, I consider the difficulty of the programmer to determine her passage through the MH event the most serious obstacle to an implementation of a relativistic computer in a Kerr-Newman spacetime. But since this is not the focus of the present paper, I assume that she can somehow determine when she perambulates the inner horizon, perhaps by applying her vast knowledge of astrophysics against the changing character of the steadily incoming radiation she receives from her fainting galactic homestead. With these assumptions in place, let us turn to the communication protocol.

\section{Communication between a computer and its programmer}
\label{sec:comm}

\subsection{The photonic communication protocol}
\label{ssec:photonic}

As we have seen, communication between the computer and the programmer is necessary for the implementation of a relativistic computer. As is obvious from the causal structure of the Kerr-Newman spacetime, this communication will necessarily be one-way from computer to programmer: the programmer cannot send a signal to the computer once she passes the outer horizon (cf.\ Figure \ref{fig:kerrnewman_penrose}). Because the programmer will then no longer be able to send instructions to the computer and its operators, an agreed-upon protocol of communication is arranged prior to the programmer's departure at $p$. In the literature, if this is specified at all, this protocol is usually based on coded photon signals. Unfortunately, this leads to various forms of the blueshift problem. First, as was noted above, the number of photon pulses sent must be limited lest the programmer be cooked to death. Thus, the above protocol according to which only a message is sent if a contradiction is found avoids this problem. 

However, there is an additional difficulty. The later the computer sends the signal, the shorter and higher-frequency the signal will be for the receiver. This will make the signal virtually impossible to be recognized or even be detected. Using various ideas in \citet{nemdav}, one can formulate three ways to solve the problem.\footnote{N\'emeti and D\'avid consider these ideas to solve distinct though related aspects of the problem. I will somewhat carelessly run these together. For further ideas of how to solve this problem, cf.\ \citet[\S4.1]{andeal09}.} First, in order for the frequencies at the receiving end to remain bounded, the computer could correct for the speed up by decreasing the frequency of the light for signals sent later. Of course, one cannot produce readable signals of arbitrarily low frequency. This impediment can be circumvented, \citet[132]{nemdav} suggest, by putting the computer on a spacecraft and send it away from the black hole of the programmer, possibly drawn by another black hole. This would effectuate a redshift in the emanated signal that could exactly counterbalance the blueshift of the original situation. While I acknowledge that this is in principle possible, this solution requires a potentially arbitrarily large amount of energy to accelerate the spacecraft or another conveniently located black hole that may not exist, as well as a perfect synchronization of the whole procedure. 

\begin{figure}
\centering
\epsfig{figure=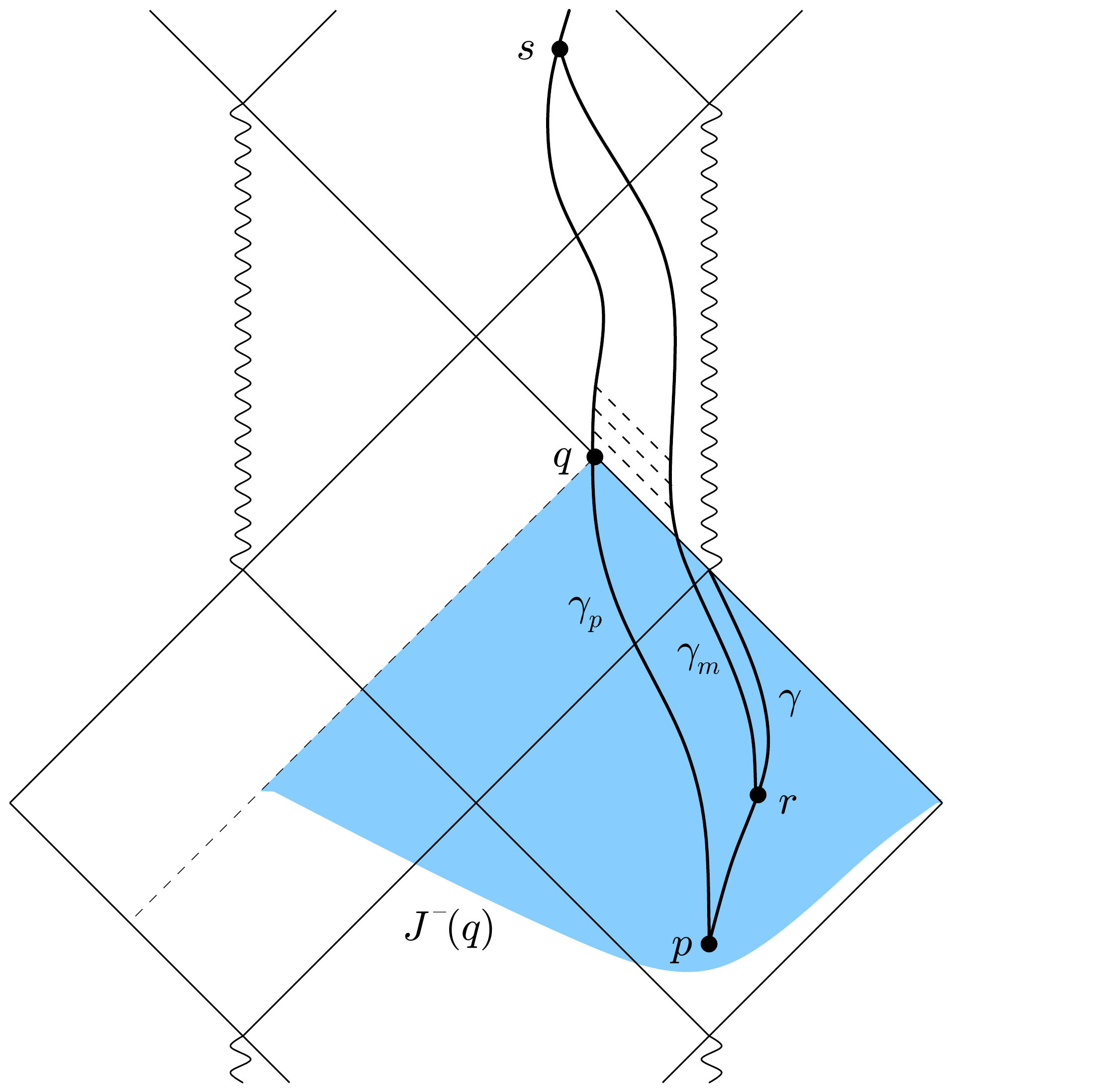,width=0.8\linewidth}
\caption{\label{fig:kerrnewman_mailman} The scenario involving a messenger travelling along $\gamma_m$ from the computer to the programmer (adapted from \citet[Fig.~5]{nemdav}.}
\end{figure}

Another solution does not rely on arbitrarily large amounts of energy or on such fine tuned features of the spacetime. Instead, the idea is to send a messenger from the computer to the programmer if and once an inconsistency of {\sf ZFC} is found. The general scenario is illustrated in Figure \ref{fig:kerrnewman_mailman}. 
So suppose the computer does find an inconsistency at $r$ and immediately hands a letter to Magyar Posta, which takes off to deliver the message to the programmer at $s$, long after the MH event at $q$. Thus, at $s$ the programmer will learn that {\sf ZFC} is inconsistent. The trouble with this protocol is that if the programmer has not received the letter, she will not know whether there is no inconsistency and thus no letter under way or whether the mail simply has not arrived yet. Thus, even if the Hungarian Post is capable of a fast physical delivery of the letter, this is not an attractive protocol. A better one would be for the mailman, whose spacecraft is roughly synchronized with the programmer's (for reasons given in \citet[\S5.3.2]{nemdav}), to start signalling the programmer once he trespasses the inner horizon, as indicated by the dashed lines in Figure \ref{fig:kerrnewman_mailman}. Although this is an improvement vis-\`a-vis the physical delivery at $s$, it is unclear how precisely this protocol can be implemented and how the programmer can be certain of the answer to the non-Turing computable question.

An altogether different solution, explicated in \citet[\S5.4.1]{nemdav}, may, if successful, not just solve the present problem but also help to stabilize the black hole against possible quantum evaporations: the computer might `feed' the black hole with matter and change the feeding patterns or change the black hole's angular momentum or its charge, in order to signal to the programmer that it found an inconsistency. The trouble with this proposal is that the acceleration of the universe needs to slow down eventually for the computer to find sufficient material supplies to continue the feeding of the black hole, for otherwise it may no longer have the means to execute the change if and when it finds an inconsistency. Even if this is so, however, the implementation of a protocol like this requires quite a bit of precision engineering! 

The challenges accompanying protocols of the above three types may be altogether avoided if what promises to be a rather elegant solution trading on non-local quantum correlations can be realized. In fact, \citet{nemdav} allude to the possibility of a protocol based on quantum non-locality:
\begin{quote}
It is an interesting future research possibility to solve the communication problem between $\gamma$ and $\gamma_p$ by some quantum-information theoretic methods. (132)
\end{quote}
This quote also shows that they do not take themselves to have fully met the challenges involved with their three proposals. And it suggests a novel vantage point to formulate a communication protocol: make use of the correlations or anti-correlations between observed properties of entangled pairs of particles. The remainder of this section is an attempt to deliver just that: the outlines of a communication protocol based on non-local quantum entanglement and to assess its prima facie feasibility. 

\subsection{A quantum-information-theoretic protocol to the rescue?}
\label{ssec:qitproto}

By way of preliminary remarks, let me explicate some background assumptions and remind the reader of the most salient features of quantum non-locality. First, I shall make the idealizing assumption that the correlations are perfect. Second, I will proceed as if it were unproblematically possible to create any number of entangled pairs of particles and to store them over an indefinite time without disturbing their state so that their entanglement persists. This is surely possible in principle, but hard in practice. Third, these correlations between observable properties of two entangled particles obtain non-locally, i.e., for spacelike separated measurement events. We will have to revisit this point below. Fourth, this `quantum connection' between entangled particles is, in the words of Tim \citet[Ch.~1]{mau02}, {\em unattenuated}, {\em discriminating}, and {\em instantaneous}---whatever that may mean in the present context. It is unattenuated because in contrast to classical (instantaneous) action, the quantum connection is unaffected by distance---making it an ideal tool for a computer of cosmic proportions. It is discriminating in that unlike gravitational forces, which affect similarly situated objects in the same manner, the quantum connection is a `private', exclusive arrangement between entangled particles. And it is instantaneous in that it seems to propagate an influence from the one particle to the other without delay---although that is a subtle point that needs to be finessed with care. 

The first stab that I am aware of at importing quantum-information theoretic protocols into the setting of Kerr-Newman spacetime was administered by \citet{xiayou}. They investigate the possibility of `quantum teleportation' in Kerr-Newman spacetime by letting one of the two parties holding an entangled particle, `Alice', hover over the event horizon of the black hole, while the other, `Bob', remains in the asymptotic region from which Alice hailed.\footnote{In keeping with my earlier gender choices, I invert the nomenclature of \citet{xiayou}.} Quantum teleportation, first proposed by \citet{beneal93}, is a quantum-information theoretic procedure to transport a quantum state in the absence of a quantum channel of communication, i.e., if only a classical channel is available.\footnote{Cf.\ \citet[\S1.3.7]{niechu} for an excellent recapitulation of quantum teleportation.} The point of the procedure is for Bob to deliver an exact copy of his one-qubit state $|\psi\rangle$ to Alice. Since the laws of quantum mechanics prevent Bob from determining the exact state of the single qubit in his possession---if he makes e.g.\ a $z$-spin measurement and finds {\sf UP}, he cannot find out whether this was a certain outcome of the system being in a $z$-spin eigenstate or merely a probabilistically determined outcome of the system being in a superposition state---, he cannot simply measure $|\psi\rangle$ and communicate it to Alice for her to obtain an exact copy. In fact, even if he knew the precise state $|\psi\rangle$, communicating it would in general take an infinite amount of classical information. So the task seems prima facie impossible.

The amazing discovery of \citet{beneal93} was that a shared entangled pair of particles permits the adoption of a protocol that solves the problem. Adapted to the present context, Alice and Bob share an entangled pair of particles before Alice travels toward the black hole, taking along her entangled particle. In brief outline, Bob lets the qubit state interact with his part of the entangled pair and then performs a measurement on both qubits to obtain one of four classical outcomes ({\sf UP-UP, UP-DOWN, DOWN-UP, DOWN-DOWN}). He then sends a message communicating the outcome of his measurement encoded in two classical bits to Alice via the classical channel. Surprisingly, if Alice now performs an operation on her part of the entangled pair depending on the message she receives, her qubit will now be transformed into an exact replica of the original qubit $|\psi\rangle$ that Bob wished to deliver to her! 

\citeauthor{xiayou} argue that one should expect a reduced fidelity in quantum teleporting states if Alice approaches the event horizon of the black hole, due to the Hawking effect. The Hawking effect is the surmised semi-classical phenomenon of particle-antiparticle pair creation in the vacuum near the event horizon such that only one of the created particle escapes the gravitational pull of the black hole to radiate into the asymptotic region, while the other falls through the horizon. This effect leads to a thermal bath of excited photons in the vicinity of the event horizon, disturbing the state of Alice's entangled particle in its cavity. The goal of their paper is to compute how high the fidelity of quantum teleportation is in the proposed scenario, using the resources of quantum field theory in curved background spacetime. Of course, this means that their considerations are at best first-order approximations that may well fail in the presence of the strong gravitational fields that our programmer may encounter. In spite of this limitation, their results may indicate a near impossibility to maintain entanglement if one of the particles is deployed to accompany the programmer through a journal into a black hole. Fortunately, they find that the fidelity is high for slow, massive black holes (as it is for extremal black holes), the situation we found above to be the most conducive to the programmer's survival.

In order to preserve the entanglement between the particles, it may also be necessary to minimize Alice's acceleration on her journey, as the amount of entanglement seems to diminish for noninertial observers \citep{fueman05}. That may not be an obstacle to our plans, as unlike in \citet{xiayou} Alice will not be brought near the event horizon and then be accelerated away from the black hole to hover over the horizon, but instead will dive through the event horizon, possibly in free fall. Both the results of \citet{xiayou} and \citet{fueman05} are semi-classical, and hence preliminary. Yet they hint at subtle, but deep, connections between gravity and entanglement that remain to be fully understood. 

The attentive reader may have noticed why quantum teleportation of the kind proposed in this recent literature will not solve the problems at hand: the protocol of quantum teleportation still requires a classical communication channel. But if the computer and our programmer could communicate classically once the computer has found an inconsistency in {\sf ZFC}, then the whole point of the present discussion of quantum-information-theoretic ways to circumvent the need for such communication would be moot! Quantum teleportation would help if the task were to transmit large amounts of information from computer to programmer, but not if it is to faithfully communicate a simple {\sf YES} or {\sf NO}. So the question before us is whether there is a way for the computer to communicate with the programmer without any `channel'. After all, at least in principle, we only need to transmit one bit of information! 

The simple idea, then, would be to implement the following protocol. Both programmer and computer each obtain one particle of an entangled pair. The programmer then departs on her journey and the computer starts its task of checking whether {\sf ZFC} is consistent. If and once the computer finds an inconsistency, then it performs a measurement or some local operation on its part of the entangled system. As long as it does not find an inconsistency, the computer leaves its particle undisturbed. The question now is whether the programmer inside the outer horizon of the Kerr-Newman black hole could somehow find out using only her entangled particle that a measurement has been made on the entangled counterpart held by the computer on the home planet.

\subsection{The limits of entanglement-based signalling}
\label{ssec:signalling}

The trouble with this simple quantum-information-theoretic protocol is that it seems to run afoul of a central theorem in quantum mechanics prohibiting `signalling'. This theorem rules out the possibility of a `Bell telephone', i.e., a way of using entangled pairs of particles in spacelike-related regions to instantaneously transfer information from one region to the other in what would constitute superluminal signalling and hence an apparent violation of special relativity. As this theorem gives the conditions under which entanglement cannot be used for signalling, we need to make sure that our protocol does not inadvertently satisfy these conditions. Roughly stated, the theorem asserts there can be no superluminal information transmission (`signalling') between spacelike-related regions even if they share an entangled pair of quantum systems. 

This theorem, or some variant of it, is usually thought to protect the integrity of relativity in the face of non-local Bell correlations. You might think that these variants of the theorem, which are all stated in terms of spacelike relations between the regions of the parts of the entangled system, do not concern the present problem, which, to repeat, is to try and use entanglement to signal between {\em timelike- or null-related} events. We cannot, however, easily evade the strictures of these results because they do not seem to make any assumption concerning the spatiotemporal separation of the subsystems. 

While there are different versions of the theorem, the following are typical premises of their proofs. First, it is assumed that the particles held by Alice and by Bob don't interact and hence that the Hilbert space of the total system is a tensor product state
\begin{equation*}
\mathcal{H} = \mathcal{H}_A \otimes \mathcal{H}_B,
\end{equation*}
where $\mathcal{H}_A$ and $\mathcal{H}_B$ are the Hilbert spaces of Alice's and of Bob's particle, respectively. Second, the state of the composite system is given by a density operator on $\mathcal{H}$, i.e., $\rho = \sum_i A_i\otimes B_i$, where $A_i$ and $B_i$ are operators on $\mathcal{H}_A$ and $\mathcal{H}_B$, respectively, which need not be states on the subsystems. A state $\rho$ that is not separable, i.e.\ cannot be expressed as $\rho = \sum_i p_i \rho_A^i \otimes \rho_B^i$, where the $\rho^i_X$ are density operators on $\mathcal{H}_X$ and $p_i \geq 0$, is called {\em entangled}. Third, it is assumed that Alice and Bob can only operate locally on their particle. Thus, the operators corresponding to the local operations that e.g.\ Alice can perform on her particle are of the form $O = O_A \otimes I_B$, where $O_A$ is some operator defined on $\mathcal{H}_A$ and $I_B$ is the identity operator on $\mathcal{H}_B$, and mutatis mutandis for Bob. It should be noted that nothing in these assumptions places specific demands on how the subsystems $A$ and $B$ are spatiotemporally related. 

Without going into the details,\footnote{Cf.\ \citet[\S2.E]{perter04} for a more rigorous statement of the results and a commented list of references for the specific results.} one can then derive that the statistics of measurement outcomes in one wing do not change under local operations in the other wing. Thus, whatever Bob does to his particle, Alice has no way of detecting Bob's local operation in her particle. In fact, she cannot even detect whether he did anything at all. 

It must be noted that in this argument, nothing regarding the measurement events' relation in time (or space) is assumed; instead, it is assumed that the local operations on a subsystem do not change the state of the other, distant subsystem. Although different variants of the no-signalling theorem differ over details, it is thus no surprise that several authors have complained that these results are relevantly circular.\footnote{Cf.\ \citet{ken95}, \citet{peahep00}, and \citet{pea09}.} In one way or another, they presuppose what they are supposed to establish, viz.\ that entangled states cannot be used to signal between what is normally assumed to be spacelike-related regions. From this it does of course not follow that such signalling is possible, let alone easy; it merely follows that it has not been shown to be impossible. While they thus fail to outlaw signalling, they show how special-relativistic constraints can be consistently implemented in the formalism of quantum mechanics. 

However, since the impossibility of superluminal signalling between space\-like-related regions can be well justified on the basis of special relativity, the assumed conditions stated in the premises of the no-signalling results arguably obtain. But the same justification does not apply to situations in which the regions are timelike- rather than spacelike-related, for there is of course no relativistic reason why signalling could not occur between these regions. Thus, `timelike signalling', as it may be called, will arguably not face the same restrictions as the spacelike case. 

If it is the case that entanglement can be `used' for signalling in all {\em and only in} situations in which we could have signalled ordinarily,\footnote{As was asserted by a referee.} then my suggestion is of course moot, as either we could have signalled anyway or else cannot use entanglement to do so. But the issue relevant for my project is whether the antecedent holds, which I take to be an open question. It seems as if in our case, ordinary signalling is not possible, and yet entanglement-based communication may be. Thus, as far as the foundations of physics are concerned, my proposal effectively rises the non-trivial and important question of the extent of the no-signalling prohibitions in quantum theory. 

All these no-signalling results, whatever their merits, are results in ordinary, i.e.\ non-relativistic, quantum mechanics augmented by the premises of these theorems, which are not axioms of ordinary quantum mechanics. Whatever else may be the case, ordinary quantum mechanics will surely not be true in strong gravitational fields. So let's move our attention to quantum field theory on curved spacetime as a first step towards a quantum theory of gravity combining quantum effects with strong gravitational fields. In fact, if we do that, there are---perhaps unsurprisingly---some indications that spacelike and timelike separations indeed differ relevantly. 

In axiomatic quantum field theory, the locality of operations is built into the axioms of `microcausality' or `locality', which assume that, roughly, operators predicated of spacelike-separated regions commute. Thus, in axiomatic quantum field theory, a `local action' principle is usually observed, where `local action' means something like `no physical action can propagate faster then the speed of light'. This can be articulated more rigorously in different ways. For instance, `Locality' in the Haag-Kastler axioms means that algebras `living' is spacelike separated regions commute, as follows:
\begin{axiom}[Haag-Kastler Locality]\label{ax:local}
Given an algebra $\mathcal{A}(\mathcal{O})$ of operators defined over a spacetime region $\mathcal{O}\subset \mathcal{M}$, with $\mathcal{O}'\subset \mathcal{M}$ denoting the set of spacetime points spacelike separated from every point in $\mathcal{O}$ and $\mathcal{A}'$ the set of operators that commute with every operator in $\mathcal{A}$ (the `commutant' of $\mathcal{A}$), then
\begin{equation*}
\mathcal{A}(\mathcal{O}') \subseteq \mathcal{A} (\mathcal{O})'.
\end{equation*}
\end{axiom}
\begin{figure}
\centering
\epsfig{figure=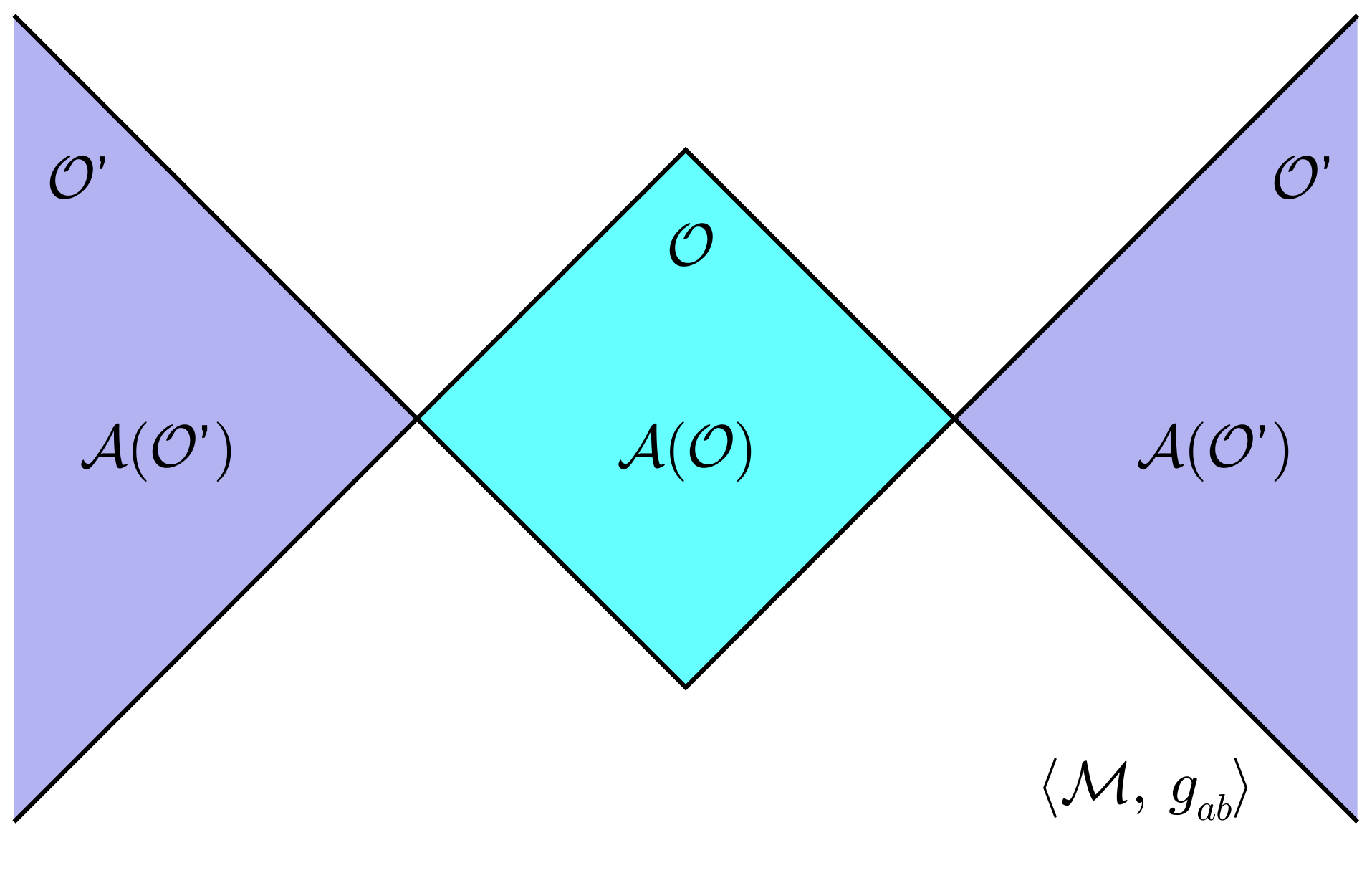,width=0.75\linewidth}
\caption{\label{fig:complement} Axiom \ref{ax:local} asserts that the algebra of the (purple) spacetime complement $\mathcal{O}'$ of a (cyan) region $\mathcal{O}$ is a subalgebra of the commutant of the algebra of $\mathcal{O}$.}
\end{figure}
In other words, operators of observables in spacelike-separated regions are among those that commute with one another. But notice what the axiom does not stipulate: that timelike-separated operators must commute. In fact, in general they will not, and so it is at least not prohibited that there can be entanglement-based timelike signalling. We should surely not consider the no-signalling theorems of standard quantum mechanics no-go theorems for a purely entanglement-based communication protocol between computer and programmer, even if they succeeded to be that for the spacelike case. However, it is also clear that such a protocol is currently no more than a mere possibility; we need a concrete protocol and a concrete understanding of how this would work.\footnote{One might worry at this point, as did a referee, that the usual conceptualization of entanglement between states in algebraic QFT required that their algebras commute and hence that the notion of timelike entanglement is meaningless for non-commuting algebras in timelike-separated spacetime regions, or---worse---that such entanglement may not exist. I cannot easily allay this concern, but only remark that if the concept of entanglement presupposes that the algebras predicated of the respective spacetime regions commute, then it would be a conceptual truth that entanglement cannot be used for signalling between these regions, quite regardless of how they are causally related. It is clear that this could not be justified on the basis of special-relativistic prohibitions of superluminal signalling alone. In what follows, I will simply assume---perhaps against better advice---that the possibility of entanglement-based `timelike signalling' remains open.}

Furthermore, it should be noted that it has recently been shown that at least for massless quantum fields in the Minkowski vacuum state there also exists field entanglement between timelike-separated regions of spacetime \citep{olsral11}. In fact, \citet[\S V]{olsral12} even speculate about a quantum teleportation protocol based on such timelike entanglement! Just as in the spacelike case above, however, their proposed protocol also relies on a classical communication channel and thus does not help our programmer. What clearly follows from my argument, though, is that whether there can be timelike signalling may be a more important question than we previously thought.

So while quantum field theory seems quite hospitable to the general possibility of timelike signalling, it turns out that on some interpretations of non-relativistic quantum mechanics, superluminal, i.e.\ spacelike, signalling is permitted.\footnote{Cf.\ the excellent survey in \citealt{ber08}, particularly \S7.} This would be possible just in case the value of some local observable controllable by Bob would influence the statistics of Alice's measurement outcomes. At least for a large class of theories, \citet[\S7.1]{ber08} argues that the following two conditions are individually necessary and jointly sufficient for superluminal signalling:
\begin{enumerate}
\item ``{\bf Controllable probabilistic dependence.} The probabilities of distant measurement outcomes depend on some nearby {\em controllable} physical quantity. 
\item ``{\bf $\lambda$-distribution.} There can be in theory an ensemble of particle pairs the states of which deviate from the quantum-equilibrium distribution; where the quantum-equilibrium distribution of pairs' states is the distribution that reproduces the predictions of orthodox quantum mechanics.''
\end{enumerate}
Since we are only looking to establish the possibility of timelike signalling, I take it that these two conditions would be jointly sufficient, though not individually necessary. 

There exist both non-collapse as well as collapse interpretation of ordinary quantum mechanics that permit superluminal signalling.\footnote{Cf.\ \S7.2 and \S7.3 of \citealt{ber08}, respectively.} First, there is Bohmian mechanics, which involves parameter independence and therefore controllable probabilistic dependence. If quantum statistics is regarded not as lawlike, but only as contingently true, then there may be perhaps atypical worlds in which $\lambda$-distribution holds. So possibly, Bohmian mechanics permits superluminal signalling. Similarly, the two conditions may obtain in some collapse theories, such as e.g.\ the so-called `non-linear continuous stochastic localization models', if we can very finely control some degrees of freedom.\footnote{Cf.\ \citet{buteal93}.} Note that in this case, the statistics would be different from orthodox quantum mechanics and we would thus have two empirically inequivalent theories. Finally, \citet{ahaeal04} proposed a signalling protocol based on the idea of `protective measurements' which supposedly permit the extraction of information from an entangled state without `disturbing' it.\footnote{Cf.\ also \citet{pea09}.}

In sum, many specific questions remain open, and will remain so until a successful quantum theory of gravity is available. But I submit that we have hitherto at least no conclusive reason to assume that quantum entanglement cannot be used for the timelike communication between the computer and the programmer behind the outer event horizon and perhaps even beyond the inner horizon. If it can be so used, it would be clear that it we would sacrifice certainty: whatever local manipulations would be performed on the computer's particle would at best alter the {\em statistics} of the measurement outcomes at the programmer's end. This has two consequences. First, the programmer would need to make multiple measurements on an ensemble of entangled particles to obtain a statistical distribution of measurement outcomes. Second, even then, only statistical confidence, but not certainty can be achieved: if the programmer performs many measurements on her particles and obtains an outcome statistics that deceptively resembles, say, the statistics we would expect if no manipulations were performed on the other end, then she will falsely come to believe that the computer did not derive the formula {\sf FALSE} and that {\sf ZFC} is thus consistent. 

The general outline of this protocol, then, would be as follows. Before the programmer leaves the computer behind to embark on her dive into the black hole, they prepare a large ensemble of entangled pairs of particles which are then individually sealed into insulating cavities. The programmer and the earthbound operator of the computer agree on a certain local operation to be performed on all particles in the terrestrial ensemble if and once the computer finds an inconsistency of {\sf ZFC}. So the programmer knows what statistics of outcomes to expect depending on whether or not such local operations have been executed. The programmer then departs for the black hole with her ensemble of particles. If and only if the computer finds an inconsistency will it manipulate its particles according to the agreed upon plan. Assuming that the programmer can determine her point of entry into the inner horizon and thus knows when she passes the MH event, she can then perform the pertinent measurements on her particles to find the resulting statistical distribution of measurement outcomes. If she does her statistics right, she can then infer, with some statistical confidence, whether or not the computer manipulated its particles and thus whether or not {\sf ZFC} is consistent.

\section{Conclusion}
\label{sec:conc}

A fair assessment of the prospects of an in-principle possible communication protocol between the computer and its programmer based on shared quantum entanglement requires a resolution of a number of issues at the very forefront of understanding quantum physics and its workings in a gravitational context. Ultimately, of course, we will need a quantum theory of gravity to accurately describe how quantum systems and strong gravitational fields interact. Only then will it be possible to close the now wide open case of how to implement a relativistic computer---or should we say a relativistic {\em quantum} computer---if that is at all physically possible. Strictly speaking, we already need to take into account quantum effects already for the purely photonic protocol: photons are quantum systems, and to consider worldlines of classical test particles which travel at the speed of light amounts to a bet that this will yield `sufficiently true' results, for which there is of course no guarantee. 

The central lesson, I take it, is that studying the physical Church-Turing thesis not only ties computational questions concerning the Church-Turing thesis with foundational issues in general relativity, such as the stability of horizons and the cosmic censorship hypothesis, but also connects with foundational problems in quantum physics, and hence with quantum gravity. In this sense, the physical Church-Turing thesis thus finds itself at the intersection of the mathematical foundations of computation, of the foundations of general relativity and cosmology, and of quantum physics. Furthermore, any penetrating analysis of relativistic computers turns on a delicate negotiation between multiple levels of theoretical considerations, e.g.\ concerning their consistency with axioms of a theory, their consonance with physical intuitions, and how insights from distinct theories ought to be combined.

Anyway, as things currently stand, I concur with the cautiously optimistic tone of \citet{nemdav}, \citet{nemand06}, and \citet*{andeal09} and conclude that the chances of such a device being at least in principle physically possible are intact. So is hypercomputation physically realizable? Well, I don't know. But we should be grateful to Istv\'an who has been so instrumental in opening up this area for fruitful research. 

\bibliographystyle{plainnat}
\bibliography{computation}

\end{document}